\documentclass[aps,prl,reprint,superscriptaddress]{revtex4-2}

\usepackage{graphicx}
\usepackage{dcolumn}
\usepackage{bm}
\usepackage[hidelinks]{hyperref}
\usepackage{xcolor}

\usepackage[normalem]{ulem}

\begin{document}

\title{Plasma wakefield dynamics of self-generated electron bunch trains}
\newcommand{\equalcontrib}{These authors contributed equally to this work.}

\author{Salome Benracassa}
\thanks{\equalcontrib}
\affiliation{Department of Complex Systems, Faculty of Physics, Weizmann Institute of Science, Rehovot, Israel}

\author{Sheroy Tata}
\thanks{\equalcontrib}
\affiliation{Department of Complex Systems, Faculty of Physics, Weizmann Institute of Science, Rehovot, Israel}

\author{Yinren Shou}
\affiliation{Department of Complex Systems, Faculty of Physics, Weizmann Institute of Science, Rehovot, Israel}

\author{Aaron Liberman}
\affiliation{Department of Complex Systems, Faculty of Physics, Weizmann Institute of Science, Rehovot, Israel}

\author{Victor Malka}
\affiliation{Department of Complex Systems, Faculty of Physics, Weizmann Institute of Science, Rehovot, Israel}

\date{\today}

\begin{abstract}
Laser plasma accelerators can deliver high-energy, quasi-monoenergetic electron beams over centimeter-scale distances. In this work, we report on the generation of narrow, quasi-monoenergetic electron bunch trains with periodic energy spacing issued from downramp injection in a laser driven wakefield accelerator. The periodicity in energy is shaped via relativistic lengthening of the wakefield during the acceleration phase, while the spatial periodicity is obtained via injection into multiple plasma periods. At the end of the accelerator, a rotation in phase-space is performed to compress each bunch in energy, producing narrow periodic spikes in the spectrum. The experimental observations are supported by particle-in-cell simulations, which reproduce the formation and evolution of the periodic bunch trains, providing an insight into the underlying plasma dynamics.

\end{abstract}
\maketitle

Laser–plasma accelerators (LPAs) can deliver high-energy electron beams in nonlinear wakefields driven by intense laser pulses, enabling acceleration over short distances through electric fields on the order of hundreds of GV/m sustained in plasma waves~\cite{Malk02}. LPAs rely on periodic plasma density modulations, within which each bucket supports alternating accelerating and decelerating fields~\cite{Tajima1979}. At relatively low laser intensities, linear plasma waves are formed, characterized by sinusoidal density variations.
As the laser intensity increases, nonlinear plasma waves develop due to modifications in the trajectories of electrons at the rear of the plasma bucket, leading to a curvature of the wavefronts~\cite{esarey1995trapping, Pukhov2002}. This behavior arises from the relativistic motion of electrons within the sheath of each plasma bubble, which effectively lengthens the plasma period and, consequently, the plasma wavelength.
In this nonlinear regime, the accelerating fields can exceed those of the linear regime by more than two orders of magnitude, while also facilitating electron injection from the background plasma and enabling the generation of stable, quasi-monoenergetic electron beams~\cite{Mangles2004, Faure2004}. Precise control of the electron injection process and of the plasma wakefield dynamics is essential for improving beam quality and expanding beam parameters.

Among the various injection techniques that enable such control, shock-front injection has proven particularly effective for generating electron beams with narrow energy spread~\cite{Bulanov1998, Geddes2008, Malka2012, chien2005, Ke2021, Ge2026}. Hydrodynamic optical-field-ionization (HOFI) shock fronts offer an additional level of control~\cite{Faure2010, Brijesh2012, Foerster2022} when compared to shocks generated with mechanical obstructions to the gas flow. In the HOFI injection scheme, a separate heater beam creates a density down-ramp through the hydrodynamic expansion of an ionized plasma column. By enabling precise control over the heater-beam profile and timing, HOFI shocks provide improved flexibility compared to shocks driven by purely hydrodynamic flows, and allow the injection position to be adjusted independently of the main laser focus~\cite{hofi_2023}.

Interestingly, under certain injection conditions, LPAs can produce electron beams with energy distributions exhibiting multiple distinct peaks. Injection schemes such as colliding-pulse injection~\cite{Faure2006, Wenz2019, Golovin2020}  and density down-ramp injection, as observed in both experiments and simulations, enable electrons to be trapped in successive plasma periods~\cite{ Golovin2020, Umstadter1996, Bourgeois2009, Gotzfried2020, Buck2013, Kurz2021, Gotzfried2020}. Simulations have also shown that injection into multiple plasma buckets can arise from laser spot-size evolution in an unmatched plasma channel~\cite{oguchi2008multiple}. More recently, multi-bunch structures have been reported using sub-cycle ionization injection driven by the evolving carrier–envelope phase of few-cycle laser pulses~\cite{Angella2025}.

We report here on the formation of a train of electron bunches periodically spaced in both energy and time. The observed energy periodicity is shaped by the non-linear evolution of the wake, where relativistic lengthening shifts the relative phase of the accelerating field for electrons injected into successive buckets, while the spatial periodicity is determined solely by the plasma wake. 
Therefore, the electron bunch train can serve as a single-shot diagnostic of relativistic plasma lengthening, offering a potential in-situ experimental probe of the non-linear wake dynamics.

\begin{figure}[t]
    \centering
    \includegraphics[width=\linewidth]{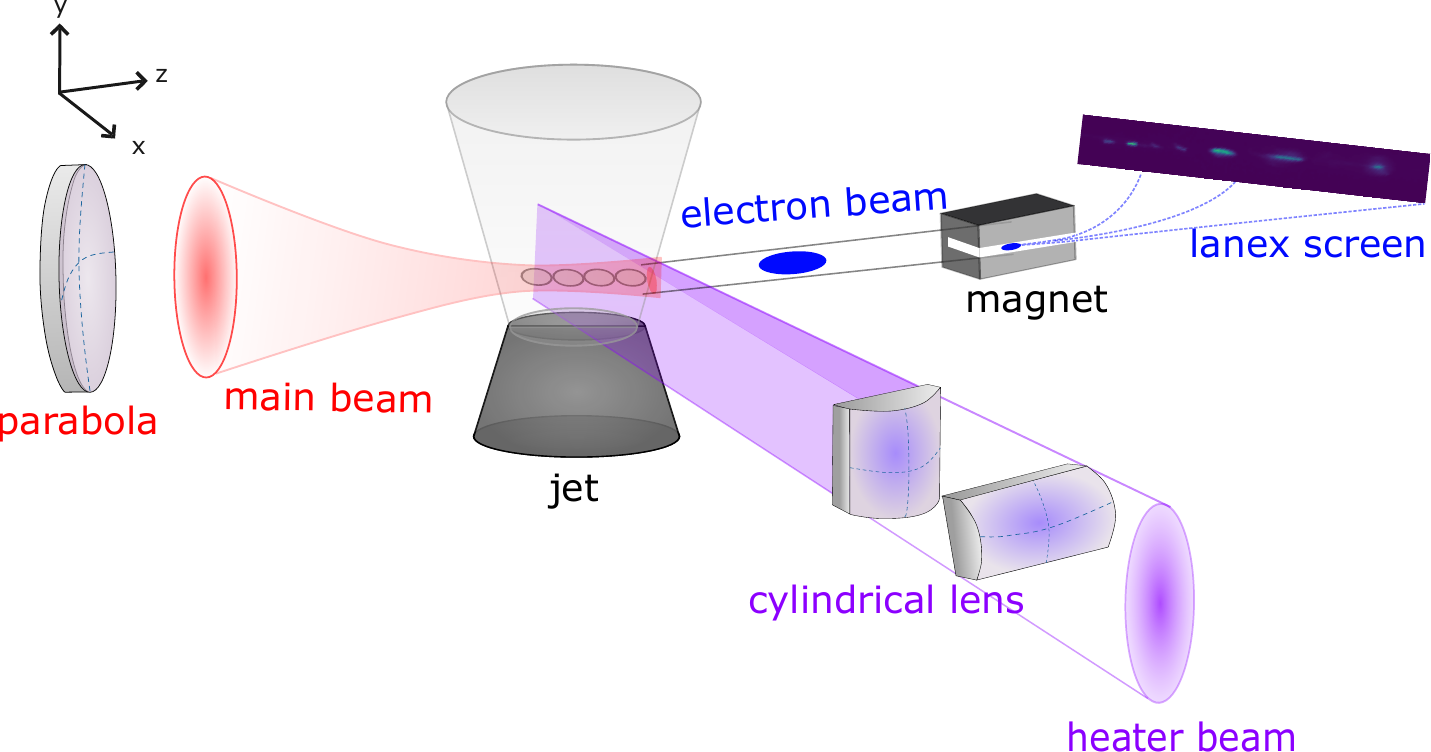}
     \caption{Schematic of the HOFI shock injection experiment setup}

    \label{fig:set_up_figure2}
\end{figure}

\begin{figure*}[t]
    \centering
    \includegraphics[width=\textwidth]{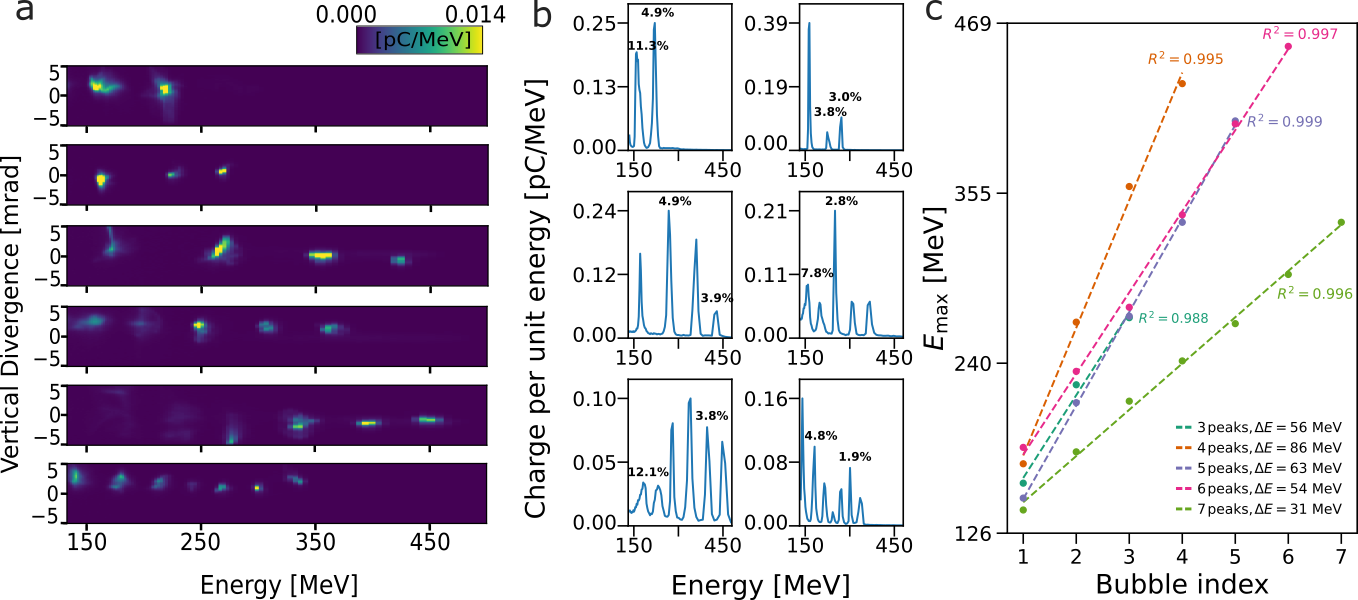}
    \caption{(a) Lanex images recorded after the magnet. The horizontal axis represents energy dispersion, and the vertical axis corresponds to beam divergence, showing shots with 2 to 7 distinct monoenergetic beamlets. (b) Corresponding energy spectra of the electron beams for the same shots, where the charge per unit energy is plotted as a function of energy. Each shot exhibits 2 to 7 separated beamlets, consistent with (a). The lowest and highest FWHM of the relative energy spread are indicated above the corresponding peaks. 
    (c) Linear fit of the peak energies extracted from (b) as a function of the beam index for all shots. Markers indicate the energies at maximum charge for each beam, and dashed lines show the linear fit. Colors represent different shots: blue (3 peaks), orange (4 peaks), purple (5 peaks), pink (6 peaks), and green (7 peaks).}
    \label{fig:data_figure} 
\end{figure*}

The experiment, illustrated in Fig.~\ref{fig:set_up_figure2}, was performed using the two beams of the HIGGINS laser system at the Weizmann Institute~\cite{kroupp2022}. The first beam served as the driver and was focused by an off-axis parabolic mirror to a peak intensity on target of $I_0 = 6 \times 10^{18}\,\mathrm{W\,cm^{-2}}$.
The second beam, oriented perpendicular to the driver in the $xy$ plane, was used for electron injection. This machining (heater) beam was focused into a line with a vertical full width at half maximum (FWHM) of $870\,\mu\mathrm{m}$ and a transverse FWHM of $3.5\,\mu\mathrm{m}$, reaching a peak intensity of $I_0 = 1 \times 10^{16}\,\mathrm{W\,cm^{-2}}$ along the line. The length of the line in the $xy$ plane was chosen to maintain a sufficiently uniform high intensity, ensuring homogeneous heating along the main beam path while providing adequate spatial overlap with the driver pulse. Along the $z$ axis (the laser propagation direction), the beam size was set to approximately one-tenth of the plasma period, creating a highly localized heating region and thus a steep density gradient. The position of this region along the laser axis was precisely controlled, enabling shock-front injection. The driver focus was scanned and positioned between $1.5\text{--}2.5\,\mathrm{mm}$ downstream of the shock, where periodic electron bunch trains were observed.
A $10\,\mathrm{mm}$ diameter supersonic gas jet was used, and the plasma density was varied between $1\text{--}2 \times 10^{18}\,\mathrm{cm^{-3}}$ by adjusting the nozzle backing pressure. This allowed optimization of the electron beam quality and stability across different parameter regimes. Pure helium was used to isolate shock-front injection and suppress continuous injection mechanisms such as ionization injection and self-injection. The delay between the driver and heater beams was varied between $0.9\,\mathrm{ns}$ and $2\,\mathrm{ns}$.

Under the experimental conditions described above, an unexpected feature emerges, the consistent formation of multiple mono-energetic peaks separated by the same energy. As shown in Fig.~\ref{fig:data_figure}(a), Lanex screens from different shots exhibit between $2$ to $7$ highly mono-energetic peaks. The corresponding processed and integrated 1D electron spectra are presented in Fig.~\ref{fig:data_figure}(b). The energy spectra exhibit periodic features, as highlighted by the linear fit of the peak energies shown in Fig.~\ref{fig:data_figure}(c).

Whereas multiple injections into successive plasma bubbles have been observed in the studies referred to above, to our knowledge, the observation of such a clean electron energy distribution, with equally spaced electron bunches, has not been experimentally observed. To elucidate the underlying mechanism, we performed Fourier-Bessel Particle-in-Cell (FBPIC) simulations~\cite{remi2016} that reveal the interplay between bunch and plasma dynamics and quantitatively reproduce the observed formation and evolution of equally spaced electron beams.

The electron density profile after the expansion of the plasma by the heater beam was computed using FLASH simulations~\cite{fryxell2000flash} based on the focused intensity profile of the heater beam. The resulting density distribution was then used as input for the FBPIC simulations. FLASH simulations predicted a density gradient of approximately $2\,\mu\mathrm{m}$, significantly shorter than the plasma wavelength of $33\,\mu\mathrm{m}$, thereby enabling localized injection into the plasma wake~\cite{Schmid2010,Suk2001}.

\begin{figure}[t]
    \centering
    \includegraphics[width=\linewidth]{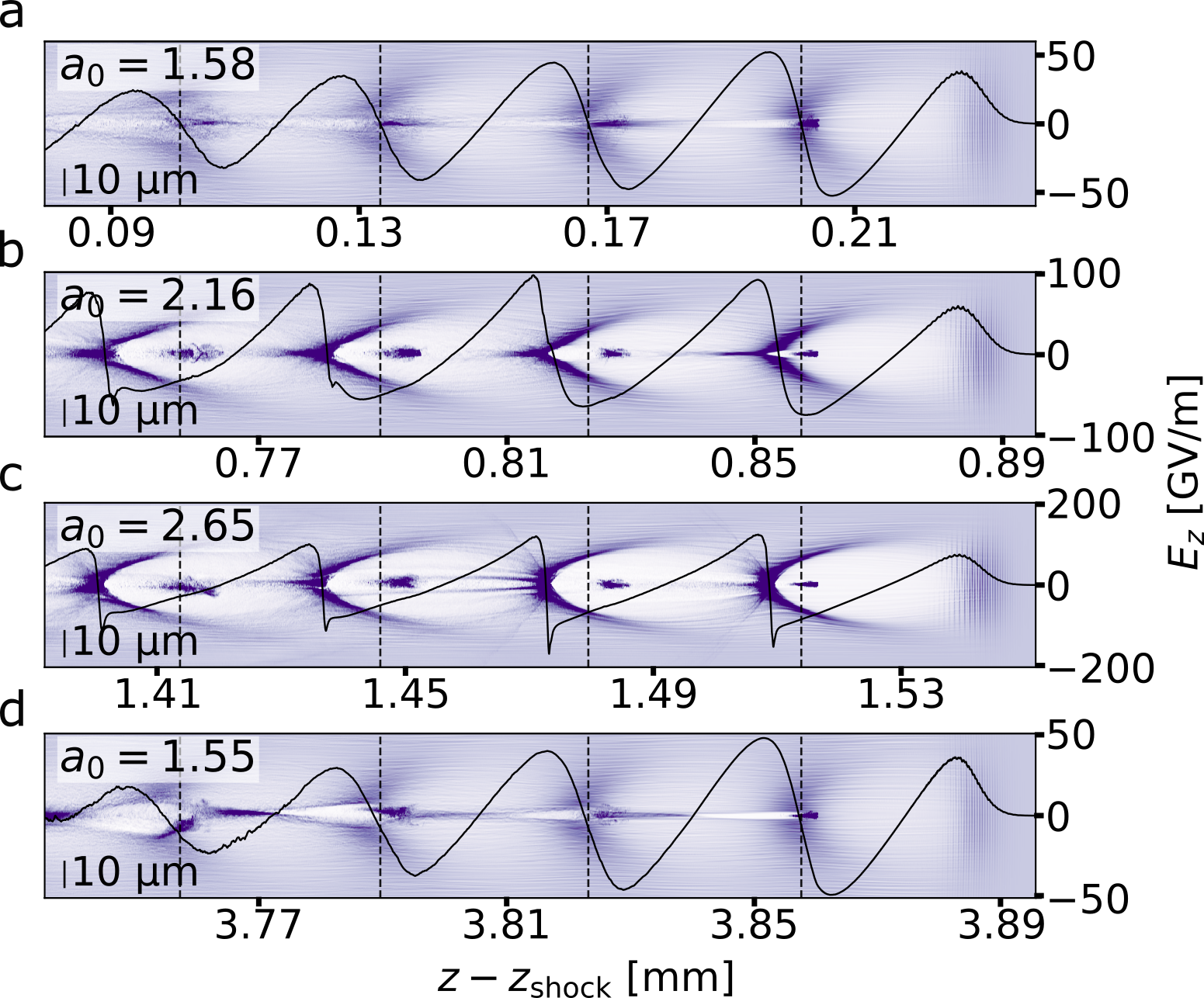} 
     \caption{Simulation of the electron density in the y–z plane at four positions downstream of the shock, with the accelerating field $E_z$ overlaid. The phase shift of injected electrons between successive plasma periods arises from relativistic lengthening driven by the evolution of the laser intensity $a_{0}$, as indicated in each panel. The shift is emphasized by dashed lines marking the relative injection positions of the electron beam within the simulation window. }

    \label{fig:electron_pos_rho_z}
\end{figure}

Figure~\ref{fig:electron_pos_rho_z} illustrates the physical origin of the periodic energy structure, which shows the electron charge density forming the wake with the injected electron beam at different positions along the gas jet. The accelerating field of the wake is depicted by the black solid line. As observed in the figure, as the wakefield evolves, electrons injected in successive plasma periods are progressively closer to the bubble center, so that electrons in later buckets experience a reduced accelerating field. In contrast, electrons injected into the first plasma period are located toward the rear of the bubble, where they remain in the strong accelerating phase of the wakefield. Overall, the electron positions exhibit a linear phase shift from one plasma period to the next.
This behavior is consistent across simulations using the experimental parameters.

To understand the origin of this linear phase shift, we note that in both the experiment and the corresponding simulations, the main laser beam was focused a few millimeters downstream of the shock, resulting in a more relativistic wake after injection.
At the injection point from the downramp generated by the HOFI shock, relativistic effects are still weak, the wake is entering the quasi-linear regime, and the accelerating field is nearly sinusoidal. At this point, electrons are injected at the back of every bubble due to the reduction of the wake phase velocity across the downramp~\cite{gons11}. 
Consequently, the positions of the electrons injected into each plasma period are determined by the local plasma density, or equivalently by the plasma wavelength, as shown in Fig.~\ref{fig:electron_pos_rho_z}(a). Therefore, electrons are initially injected at nearly the same phase near the rear of each bubble, as highlighted by the dashed lines indicating the relative injection position across the different windows.
As the laser intensity increases during focusing, the wake enters the non-linear regime and undergoes relativistic lengthening. This results in a reduction in the effective plasma frequency, which increases the plasma period, expands the bubble size, and reduces the wake-phase velocity via the relativistic shift of the plasma wavelength. 
Figure ~\ref{fig:electron_pos_rho_z}(b) shows those conditions. However, the peak field strength in successive plasma periods remains nearly unchanged, and within each bubble the accelerating field forms an approximately linear ramp.
Since the electrons are already relativistic after the injection, the backward phase shift of the wake causes electrons in successive plasma periods to be progressively deeper towards the center of the bubble, where they experience a weaker accelerating field. This evolution is illustrated in Fig.~\ref{fig:electron_pos_rho_z}(b,c). A slight decrease in the peak accelerating field from one plasma period to the next is also observed due to injected electrons, but this effect is smaller than the phase shift produced by the progressive displacement of the electrons within the wake.
As the laser propagates through the plasma, it transfers energy to the electrons and gradually defocuses, reducing its intensity. Consequently, the wakefield becomes less relativistic, the effective plasma frequency increases again, and the wakefield amplitude decreases. Near the end of the interaction, as shown in Fig.~\ref{fig:electron_pos_rho_z}(d), the wakefield returns toward its non-relativistic period, shifting the phase of the electrons back toward the rear of the bubble, close to their original injection phase.
It is important to note that most of the energy gain in the wake occurs during the relativistic lengthening phase of the plasma period. The mean energy gained in each bubble is therefore representative of the accelerating potential produced by the relativistic lengthening of the plasma wake.

\begin{figure}[b]
    \centering
    \includegraphics[width=\linewidth]{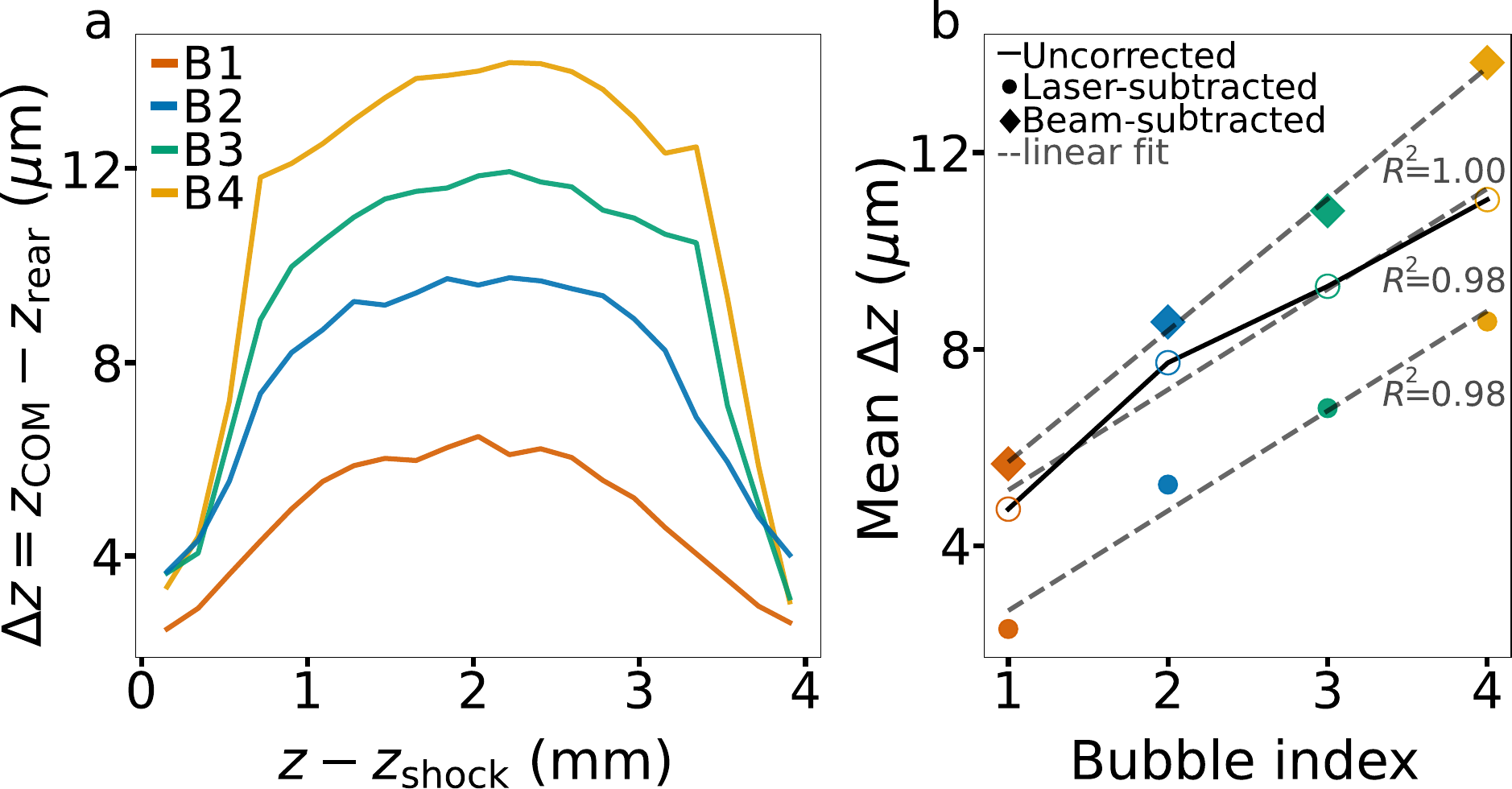} 
     \caption{Simulation evidence of linear phase slippage of injected electrons within the plasma wake.
(a) Electron beam center of mass relative to the bubble rear $\Delta z$  versus propagation distance $z$, for successive plasma periods (red: closest to the laser; yellow: farthest).
(b) Mean $\Delta z$ versus plasma period, showing the raw data (solid lines), data referenced to the laser centroid motion (squares), and data referenced to the beam COM motion arising from transverse betatron oscillations (circles).}
\label{fig:COM_vs_bubble_PRL_2rows_} 
\end{figure}

Figure~\ref{fig:COM_vs_bubble_PRL_2rows_}(a), obtained from the same simulation as Fig.~\ref{fig:electron_pos_rho_z}, shows the distance $\Delta z$ between the electron-beam center of mass and the rear of the bubble as a function of propagation distance for four successive plasma periods (B1-B4). Electrons are injected near the rear of the wake at the shock ($z-z_{\mathrm{shock}}=0$) and undergo a progressive phase shift that reaches a maximum at the laser focus ($z - z_{\mathrm{shock}} = 2\,\mathrm{mm}$), before decreasing again as the laser defocuses. The figure shows that the phase shift of the injected electrons increases with the plasma period in which they are trapped, as discussed above.
This behavior can be understood from the expansion of the wakefield, causing each plasma period to expand by approximately the same amount. Since the electrons are already relativistic, they cannot slip backward within the wake, and the expansion leads to a cumulative longitudinal shift: electrons injected into successive plasma periods are displaced by multiples of a constant offset $\Delta z$. As a result, they occupy progressively linear shifted phases of the approximately linear accelerating field, causing electrons in later buckets to experience weaker fields and leading to the observed periodic energy spacing.
A larger lengthening results in a greater energy spacing between successive bubbles.
We further observe that excessive relativistic lengthening can lead to electron loss from later plasma periods due to deceleration or during subsequent bubble contraction as the laser intensity decreases from defocusing.

Figure~\ref{fig:COM_vs_bubble_PRL_2rows_}(b) confirms that the linear shift in $\Delta z$ does not arise from laser motion or beam dynamics. In this panel, $\Delta z$ is averaged over each bubble index (B1–B4). The circles denote data in the laser-referenced frame, accounting for laser dephasing induced by density variations, while diamonds denote data with the beam centroid motion arising from transverse betatron oscillations removed. The linear dependence of $\Delta z$ on plasma period remains. This supports that the phase shift is governed by the evolving wake structure, with relativistic lengthening identified as the physical mechanism responsible for the periodic energy structure observed in both experimental data and simulations. 

\begin{figure}[h]
    \centering
    \includegraphics[width=1\linewidth]{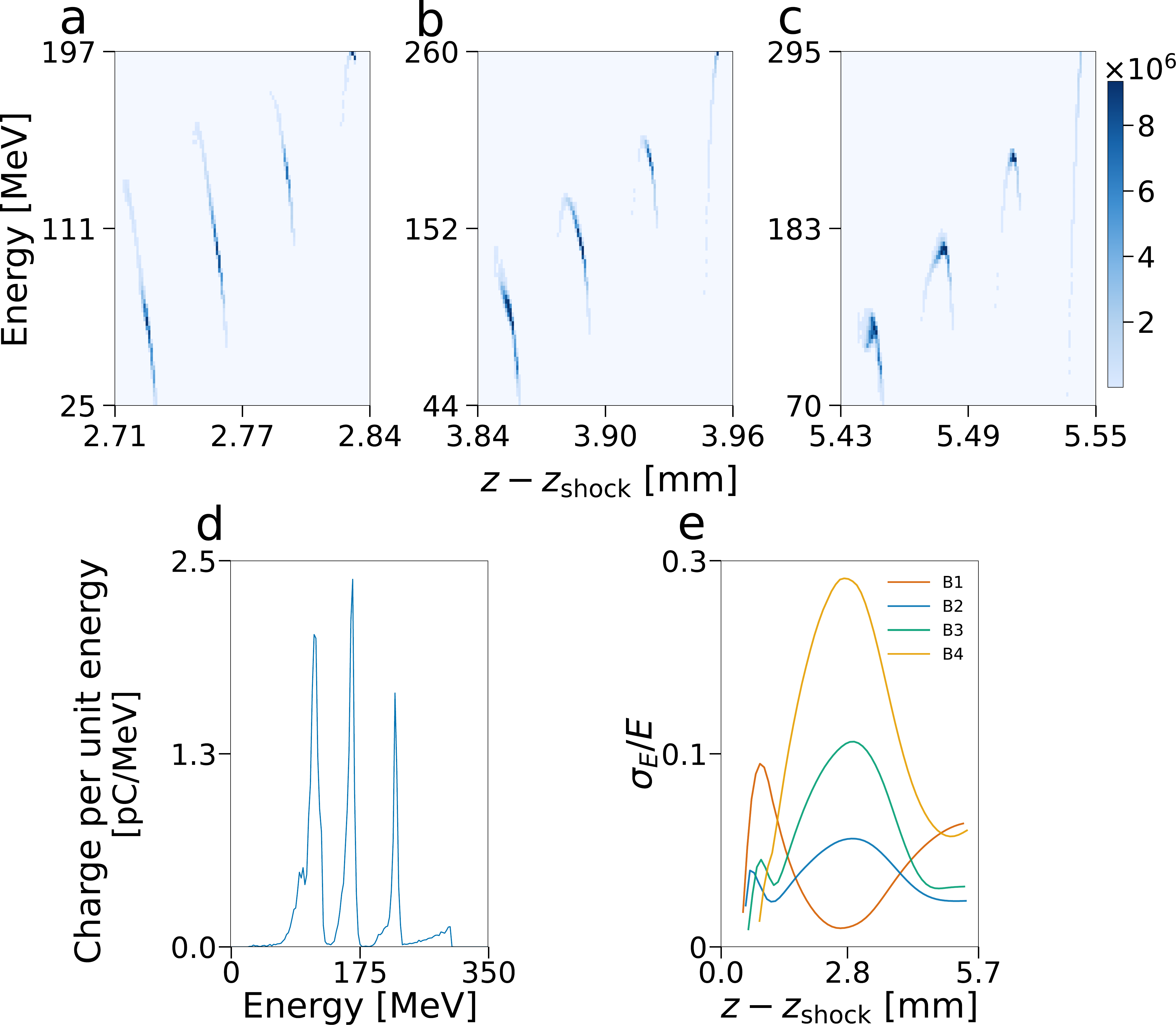}
    \caption{(a–c) Energy as a function of $z$ for a simulation window at three different positions along the plasma propagation. Electron bunches injected in successive plasma periods are clearly distinguished: the rightmost bunch corresponds to the first bubble behind the laser, while the leftmost corresponds to the fourth bubble. (d) Corresponding energy spectra at the end of the simulation.
    (e) Relative RMS energy spread as a function of plasma propagation.
    }
    \label{fig:rms_spectrum}
\end{figure}

While relativistic lengthening determines the relative phase of the electron bunches in the wake and thus explains the periodicity in energy, it does not by itself account for the formation of narrow energy peaks. As shown in Fig.~\ref{fig:rms_spectrum}, monoenergetic bunches arise from an additional process, referred to as energy compression, which is driven by the evolution of the electron phase space within the wakefield after acceleration. This energy compression results from a phase-space rotation in which lower-energy electrons at the front of the bunch gain energy at the cost of energy from the tail and approach the energy of higher-energy electrons at the back~\cite{dopp2018}.  
This process takes place near the transition between the laser-driven and beam-driven regimes. In the laser-driven regime shown in Fig.~\ref{fig:rms_spectrum}(a), electrons at the back of the bunch experience a stronger accelerating field than those closer to the bubble center. As the laser defocuses and beam loading becomes significant, the field structure starts to reverse, as shown in Fig.~\ref{fig:rms_spectrum}(b), and electrons at the back of the bunch begin to decelerate relative to those at the front~\cite{Kirchen2021,rechatin2009}. The competition between these two effects leads to a temporary reduction of the energy spread until compression is achieved, as illustrated in Fig.~\ref{fig:rms_spectrum}(c). This results in the conservation of the mean energy of the beam, enabling the probe of the mean accelerating potential. It is important to note that beam acceleration and energy compression can occur simultaneously, while simulations show that the energy gain is minimal during compression.  Without this energy compression, the final spectrum would remain continuous, since the energy distributions of the different bunches overlap before compression, as shown in Fig.~\ref{fig:rms_spectrum}(a-b).

Strong monoenergetic spectra are obtained only when compression occurs just before the electrons exit the gas jet, as seen in Fig.~\ref{fig:rms_spectrum}(d). If beam loading begins too early or the interaction length is not optimal, the bunch either over-compresses or fails to compress, resulting in a broad spectrum. In the simulations, shifts of only $0.5\,\mathrm{mm}$ in the focus of the accelerating beam were sufficient to significantly modify the compression, highlighting the sensitivity of this regime to the experimental parameters. Figure~\ref{fig:rms_spectrum}(d) shows the corresponding output spectra, while (e) shows the evolution of the energy spread as a function of propagation distance z. The energy spread reaches a maximum when the laser begins to defocus, and beam loading becomes significant, and then decreases until optimal compression is achieved at $z - z_{\mathrm{shock}} \approx 5~\mathrm{mm}$ after the shock, near the end of the gas jet. At this stage, the output spectrum is compressed into multiple monoenergetic peaks.

In most laser–plasma accelerator (LPA) simulations, only the first few plasma periods are modeled to reduce computational cost, which may explain why multi-bucket injection has rarely been reported. Our results suggest that injection into multiple plasma periods should be common as long as the density gradient is sufficiently steep.
For shock injection to obtain monoenergetic electron beams, the laser is typically focused near the shock, where its intensity is highest, and the plasma period is already elongated due to relativistic lengthening at the injection point, and achieving injection into multiple plasma periods. Our simulations, however, show that electrons injected into later plasma periods are subsequently lost as the laser weakens after the injection point, the plasma bubbles contract towards their non-relativistic plasma period, causing the bubbles to overtake the electrons. The sheath of the electrons of the plasma wake crossing the injected electrons causes the beam to be lost. As a result, only the beams injected into the first one or two plasma periods remain to form the observed output spectrum.
The periodic bunch train in both space and energy enables the beam to be used for a range of applications. The periodic spacing between bunches can lead to coherent radiation at the plasma frequency and its harmonics, enhancing THz or multi-bunch x-ray emission in an undulator or plasma radiator. 
The bunch-train structure also enables beam-driven wakefield schemes in which one bunch acts as the driver and subsequent bunches serve as witnesses. 
In addition, inverse Compton scattering of a laser from individual bunches could provide a self-referenced measurement of the electron energy loss in a single shot. 

\begin{acknowledgments}
The research was supported by the Schwartz/Reisman Center for Intense Laser Physics, the Benoziyo Endowment Fund for the Advancement of Science, the Israel Science Foundation, the Wolfson Foundation, the Schilling Foundation, and R. Lapon, Dita, and Yehuda Bronicki.
\end{acknowledgments}

\bibliographystyle{apsrev4-2}  
\bibliography{references_arxiv} 
\end{document}